# GIANT RADIATIVE INTERACTIONS AMONG DISTANT ATOMS

**G S Agarwal[1] and S Dutta Gupta[2]**

1. Physical Research Laboratory, Navrangpura, Ahmedabad-380009. INDIA.
2. School of Physics, University of Hyderabad, Hyderabad-500046. INDIA.

*We examine the feasibility of enhancing the fundamental radiative interactions between distant atoms. We present general arguments for producing enhancement. In particular, we show how giant dipole-dipole interaction can be produced by considering dipoles placed close to micron sized silica spheres. The giant interaction arises as the whispering gallery modes can resonantly couple the dipoles.*

**Introduction**

It is well known from the early days of quantum electrodynamics[1] that the exchange of a photon between two atoms leads to the radiative coupling between the atoms. It is also known that the strength of such a radiative coupling is significant only if the distance between the two atoms is much smaller than the wavelength of the transition involved in the coupling.. In this work we demonstrate the possibility of remarkably large radiative coupling between atoms which could be separated by distances much larger than the wavelength. In what follows, we concentrate on one specific type of the radiative coupling, viz., the retarded dipole-dipole interaction. We begin by recalling how the radiative coupling arises.

Consider, for example, two atoms $A$ and $B$ located at positions $\vec{r}_A$ and $\vec{r}_B$, respectively. Let at time $t=0$ the atom $A$ be in the excited state $|e_A\rangle$, and the atom $B$ in the ground state $|g_B\rangle$. Such a selective excitation can be achieved by using an optical fiber tip. We assume an allowed dipole transition between $|e_A\rangle$ and $|g_A\rangle$. Clearly we can have a transfer of excitation from the atom $A$ to $B$, i.e., the initial state of the system has changed from $|e_A, g_B\rangle$ to $|g_A, e_B\rangle$. The electromagnetic field is in vacuum state at $t=0$ and is again in vacuum state after the transfer of excitation is completed. This process $|e_A, g_B\rangle$ to $|g_A, e_B\rangle$ is mediated by the vacuum of the electromagnetic field. This transfer process can also be described by an effective interaction, which depends only on the atomic degrees of freedom. Thus, the atoms $A$ and $B$ get coupled via the radiative interaction. The amplitude for such a process can be calculated in the framework of quantum electrodynamics using the second order perturbation theory. Let $\vec{d}_\alpha$ be the dipole matrix element for the transition $|e_\alpha\rangle \rightarrow |g_\alpha\rangle$. As shown in the appendix a quantum electrodynamic calculation shows that the effective interaction is related to the dyadic Green's function $G_{ij}(\vec{r}_A,\vec{r}_B,\omega)$ associated with the wave equation for the electromagnetic field. The dyadic Green's function is related to the field $E_i(\vec{r}_A,\omega)$ produced by an oscillating dipole at the position $\vec{r}_B$

$$E_i(\vec{r}_A,\omega)=\frac{\omega^2}{c^2}\sum_j G_{ij}(\vec{r}_A,\vec{r}_B,\omega)\,p_j\ . \tag{1}$$

The function $G$ is well known to be[2]

$$G_{ij}(\vec{r},\vec{r}',\omega)=\left(\delta_{ij}+\frac{c^2}{\omega^2}\frac{\partial^2}{\partial r_i\partial r_j}\right)\frac{\exp\left(i\frac{\omega}{c}|\vec{r}-\vec{r}'|\right)}{|\vec{r}-\vec{r}'|}. \tag{2}$$

The role of G in determining the radiative interactions can be easily appreciated by examining the classical equation of motion for a dipole at the position $\vec{r}_A$

$$\left(\frac{d^2}{dt^2}+\omega_0^2\right)\vec{p}_A(t)=\frac{e^2}{m}\vec{E}(\vec{r}_A,t). \tag{3}$$

Here $\vec{E}(\vec{r}_A,t)$ is the net field at location $A$ and it should also include the self-fields[3], i.e., its Fourier component is given by

$$\vec{E}(\vec{r}_A,\omega)=\frac{\omega^2}{c^2}\left[\overleftrightarrow{G}(\vec{r}_A,\vec{r}_A,\omega).\vec{p}_A+\overleftrightarrow{G}(\vec{r}_A,\vec{r}_B,\omega).\vec{p}_B\right]. \tag{4}$$

We will now make slowly varying envelope approximation

$$\vec{p}_\alpha(t)=p_\alpha\vec{n}_\alpha e^{-i\omega_0 t}+c.c., \tag{5}$$

where $\vec{n}_\alpha$ represents the orientation of the dipole at $\vec{r}_\alpha$. We can then obtain the reduced equation for $p$

$$\dot{p}_A=\frac{ie^2\omega_0}{2mc^2}\begin{bmatrix}\vec{n}_A\cdot\overleftrightarrow{G}(\vec{r}_A,\vec{r}_A,\omega_0)\cdot\vec{n}_A p_A\\ +\vec{n}_A\cdot\overleftrightarrow{G}(\vec{r}_A,\vec{r}_B,\omega_0)\cdot\vec{n}_B p_B\end{bmatrix}\equiv -iv_A p_A-iv_{AB}p_B\ . \tag{6}$$

Thus the net complex coupling $v_{AB}$ between two dipoles is given by

$$v_{AB}=-\frac{e^2\omega_0}{2mc^2}\vec{n}_A\cdot\overleftrightarrow{G}(\vec{r}_A,\vec{r}_B,\omega_0)\cdot\vec{n}_B\ . \tag{7}$$

The real and imaginary parts of the Green's tensor determine the strength of the radiative interaction. The radiative interaction is an oscillatory function of $(\vec{r}_A-\vec{r}_B)$ and is especially important if $\frac{\omega_0}{c}|\vec{r}_A-\vec{r}_B|<<2\pi$. If the distance between the two dipoles is much larger than a wavelength, then the radiative coupling is negligibly small. The question that we ask is – is it possible to enhance the radiative coupling between atoms which are separated by a distance that is many times the wavelength of the atomic transition. As noted above the radiative interaction arises from the interaction with the vacuum of the electromagnetic field. Thus we should be able to enhance $v$ by suitably modifying the vacuum. In other words, if the structure of $\overleftrightarrow{G}$ for the modified vacuum were such that $\overleftrightarrow{G}$ had a resonance at frequency corresponding to the atomic transition frequency, then $v$ can be enhanced by a large factor. The actual value of the enhancement factor will depend on the widths of the resonances in $\overset{\leftrightarrow}{G}$. There have been several studies on surface enhanced dipole-dipole interaction [4,5]. A situation which is particularly attractive is shown in Fig.1., where we place two atoms near the surface of a micro-sphere with dielectric constant $\varepsilon$. It is known that micro-spheres support whispering gallery modes (WGM) and some of these modes with frequency say $\omega_G$ can have ultra-narrow width. In other words, if $\omega_0$ is in the vicinity of $\omega_G$, then $\overset{\leftrightarrow}{G}$ has primarily a resonant character and a small background. In what follows, we demonstrate a very large enhancement of the radiative interaction between two atoms separated by a distance of, say, two orders of magnitude larger than the wavelength of transition. Note that in recent times whispering gallery modes [6] (WGM) have attracted considerable attention following the work of Braginsky et al [7] who showed the possibility of WGM's with quality factors $Q \geq 10^8$, effective quantization volume of the order of $10^{-9}$ $cm^3$. These two properties make WGM's specially attractive in many different applications in nonlinear optics [8] and in quantum electrodynamics [9-15]. Braginsky et al had already reported the possibility of extremely low threshold for optical bistability. Haroche and coworkers have extensively studied the properties of WGM's [12]. These authors even developed newer techniques for studying such modes. Arnold and coworkers [13] studied WGM induced enhancement of energy transfer, albeit for smaller spheres (of size $\approx 5\mu m$), though as will be seen in our work that much larger enhancement can be achieved by using larger spheres. Kimble et al [14] and the present authors [15] considered interaction of WGM's with atoms. In particular, Dutta Gupta and Agarwal [15] considered micron sized spheres doped uniformly with resonant atoms. They demonstrated how effects like vacuum field Rabi splittings can be studied using WGM's. The lasing characteristics of doped micro-spheres have also been studied extensively [16].

**Structure of $\overset{\leftrightarrow}{G}$**

In order to obtain $\overset{\leftrightarrow}{G}$ for the situation shown in Fig.1, we have to solve Maxwell's equations subject to the boundary conditions at the surface of the sphere. The Greens function consists of two contributions [11] – (a) $\overset{\leftrightarrow}{G}^{(0)}$ - which is the same as in free space, i.e., in the absence of the micro-sphere; (b) $\overset{\leftrightarrow}{G}^{(s)}$ - which depends on the presence of the micro-sphere. The basic equations for these Green's functions are

$$\vec{\nabla}\times\vec{\nabla}\times\overset{\leftrightarrow}{G}^{(0)}(\vec{r},\vec{r}\,',\omega)-\frac{\omega^2}{c^2}\varepsilon_0\overset{\leftrightarrow}{G}^{(0)}(\vec{r},\vec{r}\,',\omega)=4\pi\overset{\leftrightarrow}{I}\delta(\vec{r}-\vec{r}\,'), \tag{8}$$

$$\vec{\nabla}\times\vec{\nabla}\times\overset{\leftrightarrow}{G}^{(s)}(\vec{r},\vec{r}\,',\omega)-\frac{\omega^2}{c^2}\varepsilon_0\overset{\leftrightarrow}{G}^{(s)}(\vec{r},\vec{r}\,',\omega)=0 \; . \tag{9}$$

These Green's functions are given in terms of the vector spherical harmonics as

$$\vec{G}^{(0)}(\vec{r}_A,\vec{r}_B,\omega)=ik_0\sum_{n=1}^{\infty}\sum_m(2-\delta_{m,0})\frac{(2n+1)(n-m)!}{n(n+1)(n+m)!}$$
$$\times[\vec{M}^{(1)}_{\binom{e}{o}mn}(k_0\vec{r}_>)\ \vec{M}_{\binom{e}{o}mn}(k_0\vec{r}_<)\qquad(10)$$
$$+\vec{N}^{(1)}_{\binom{e}{o}mn}(k_0\vec{r}_>)\ \vec{N}_{\binom{e}{o}mn}(k_0\vec{r}_<)],$$

$$\vec{G}^{(s)}(\vec{r}_A,\vec{r}_B,\omega)=ik_0\sum_{n=1}^{\infty}\sum_m(2-\delta_{m,0})\frac{(2n+1)(n-m)!}{n(n+1)(n+m)!}$$
$$\times[\vec{M}^{(1)}_{\binom{e}{o}mn}(k_0\vec{r}_B)\ \vec{M}^{(1)}_{\binom{e}{o}mn}(k_0\vec{r}_A)A_{\binom{e}{o}n}\qquad(11)$$
$$+\vec{N}^{(1)}_{\binom{e}{o}mn}(k_0\vec{r}_B)\ \vec{N}^{(1)}_{\binom{e}{o}mn}(k_0\vec{r}_A)B_{\binom{e}{o}n}].$$

In Eqs.(10) and (11) $\vec{r}_>$ ($\vec{r}_<$) represent the greater (smaller) of $\vec{r}_A$ and $\vec{r}_B$, $k_0=\omega/c$. The vector spherical harmonics $\vec{M}$ and $\vec{N}$ are given by [17]

$$\vec{M}_{\binom{e}{o}mn}=\mp\frac{m}{\sin\theta}j_n(kr)P_n^m(\cos\theta)\binom{\sin}{\cos}m\phi\hat{\theta}-j_n(kr)\frac{\partial P_n^m}{\partial\theta}\binom{\cos}{\sin}m\phi\hat{\phi},$$

$$\vec{N}_{\binom{e}{o}mn}=\frac{n(n+1)}{kr}j_n(kr)P_n^m(\cos\theta)\binom{\cos}{\sin}m\phi\hat{r}$$
$$+\frac{1}{kr}[krj_n(kr)]'\left[\frac{\partial P_n^m}{\partial\theta}\binom{\cos}{\sin}m\phi\hat{\theta}\mp\frac{m}{\sin\theta}P_n^m(\cos\theta)\binom{\sin}{\cos}m\phi\hat{\phi}\right].\qquad(12)$$

The vector functions $\vec{M}^{(1)}$ and $\vec{N}^{(1)}$ are obtained from Eq.(12) by replacing $j_n$ by $h_n^{(1)}$. *A* and *B* in Eq.(11) are determined by the following expressions

$$A_n=\frac{j_n(ka)[k_0aj_n(k_0a)]'-j_n(k_0a)[kaj_n(ka)]'}{h_n^{(1)}(k_0a)[kaj_n(ka)]'-j_n(ka)[k_0ah_n^{(1)}(k_0a)]'},\qquad(13)$$

$$B_n=\frac{\varepsilon j_n(ka)[k_0aj_n(k_0a)]'-j_n(k_0a)[kaj_n(ka)]'}{h_n^{(1)}(k_0a)[kaj_n(ka)]'-\varepsilon j_n(ka)[k_0ah_n^{(1)}(k_0a)]'}.\qquad(14)$$

In (13) and (14) $k=k_0\sqrt{\varepsilon}$. The vanishing of the denominators in (13) and (14) yields the various modes of the system

$$h_n^{(1)}(k_0a)[kaj_n(ka)]'-j_n(ka)[k_0ah_n^{(1)}(k_0a)]'=0,\qquad(15)$$

$$h_n^{(1)}(k_0a)[kaj_n(ka)]'-\varepsilon j_n(ka)[k_0ah_n^{(1)}(k_0a)]'=0.\qquad(16)$$

Depending on whether Eq(15) or (16) is satisfied, the modes are labeled as a-(TM) or b- (TE), respectively, since the radial component of the magnetic or electric field is absent in the corresponding field distribution. The roots of Eqs.(15) or (16) and hence the modes can be labeled by two integers *n* and *l*, where *n* gives the mode number and *l* gives the order number. Physically *l* determines the number of extrema in the radial distribution of the field. The smaller the value of *l*, the more the field is concentrated near the inner edge of the sphere. In the ray picture this corresponds to a ray, confined near the inner edge, undergoing multiple total internal reflections from inside the micro-sphere, justifying the name “whispering gallery mode”. Under plane wave illumination (as in extinction studies) a variely of such modes can be excited. Especially interesting are the modes of larger spheres with large *n*, for which the corresponding quality factors can achieve dramatic values (order of $10^{9-10}$ in the visible domain). Note that it is extremely difficult to achieve such figures with conventional Fabry-Perot cavities.
It is now clear that at chosen discrete set of mode frequencies satisfying Eqs.(15), (16), the poles in $A$ or $B$ can lead to dramatic enhancements in $\vec{G}^{(s)}$ and thus in radiative interactions. In order to demonstrate the enhancement and to facilitate a comparison of small and large spheres to reveal the potentials of the larger system, we picked silica spheres of two radii, namely, 200 μm and 5 μm. For the larger (smaller) sphere out of the variety of modes we selected $a_{2312,167}$ ($a_{39,1}$) mode with frequency $\omega_G$ = 1.8026933000486 $\mu m^{-1}$ (1.096835618 $\mu m^{-1}$) and quality factor Q = $1.28 \times 10^9$ ($1.22 \times 10^3$). In the next Section we define the quantities of interest and present numerical results pertaining to these modes.

## Giant Enhancement in the Radiative Interaction

We now discuss the relevant quantities capturing the giant enhancement in radiative interactions. We also note the presence of the self-terms in the equation of motion for the dipoles. Such self-terms give rise to the usual lifetimes and frequency shifts. These are also affected by the presence of the micro-sphere and thus, in what follows, we give a brief discussion of these quantities.
Since the free space characteristics of coupled dipoles are well known, we concentrate on the effects produced by the micro-sphere. We normalize all frequencies to the free space decay rate denoted by $2\gamma_0$ with

$$\gamma_0 = \frac{e^2\omega}{2mc^2}\,\mathrm{Im}\left[\vec{n}_A \cdot \vec{G}^{(0)}(\vec{r}_A,\vec{r}_A,\omega)\cdot \vec{n}_A\right] = \frac{e^2\omega^2}{3mc^2}. \tag{17}$$

Thus quantities of interest will be

(a) Micro-sphere dependent width $\gamma_0 K_s$ and shift $\gamma_0 \Omega_s$ of the dipole’s frequency

$$\begin{aligned}\gamma_0(K_s + i\Omega_s) &= -\frac{ie^2\omega}{2mc^2}\vec{n}_A \cdot \vec{G}^{(s)}(\vec{r}_A,\vec{r}_A,\omega)\cdot \vec{n}_A \\ &= \sum_{n=1}^{\infty} n(n+1)(2n+1)\frac{[h_n^{(1)}(k_0 r_A)]^2}{k_0^2 r_A^2}B_n ,\end{aligned} \tag{18}$$

(b) the radiative interaction

$$\begin{aligned}\gamma_0(K_d + i\Omega_d) &= -\frac{ie^2\omega}{2mc^2}\vec{n}_A \cdot \vec{G}^{(s)}(\vec{r}_A,\vec{r}_B,\omega)\cdot \vec{n}_B \\ &= \sum_{n=1}^{\infty} (-1)^n n(n+1)(2n+1)\frac{h_n^{(1)}(k_0 r_A)h_n^{(1)}(k_0 r_B)}{k_0^2 r_A r_B}B_n .\end{aligned} \tag{19}$$

The last expressions of (18) and (19) follow from Eq.(11) for radially oriented dipoles on z-axis and include the contributions from all possible modes.

In what follows, we demonstrate giant enhancement in characteristics like $\Omega_d$ due to the excitation of the whispering gallery modes. The modes associated with (16) are especially important. We begin by examining the extinction coefficient $Q_{\text{ext}}$, which is a measure of the on-axis transmission for plane wave illumination. In terms of the coefficients $A_n$ and $B_n$ given by Eqs.(13), (14), $Q_{\text{ext}}$ is defined as

$$Q_{ext} = -\frac{2}{k^2 a^2} \sum_{n=1}^{\infty} (2n+1)\,\mathrm{Re}(A_n + B_n)\,. \qquad (20)$$

The results of numerical computations for the $a_{2312,167}$ mode are shown in Fig. 2. In order to assess the contribution of the WGM we have shown in Fig.2a the extinction coefficient $Q_{\text{ext}}$ with (solid line) and without (dashed line) the resonant term. The contribution of the WGM $a_{2312,167}$ can easily be appreciated by noting the narrow (note the scale in the horizontal axis) resonance at $\omega_G$. Moreover, the background has relatively insignificant contribution to $Q_{\text{ext}}$. The results for $K_d$ and $\Omega_d$ for equidistant dipoles ($d_1 = d_2 = d$) and for two normalized distances, namely, $d/\lambda = 0.1$ and 1.0 are shown in Fig.2b. The effect of increasing distance of the dipoles from the surface is clear from a significant reduction of the enhancement factor $K_{0d} = \max(|K_d|)$ from 1349.81 to 154.47. Note that single dipole enhancement factor $K_{0s}$ has the same magnitude for the configuration under study. Analogous results for $a_{39,1}$ mode are shown in Fig.3. In contrast to the case of $a_{2312,167}$ mode, now the background plays an important role. Moreover, the enhancement factors for the same distances, i.e., $d/\lambda = 0.1$ and 1.0, are now reduced to 3.4 and 0.07, respectively. The distortion of the curves in Fig 3b is due to a nearby WGM $a_{44,1}$. We now turn to a direct comparison of the enhancement factors $K_{0d}$ as functions of the distance of the second dipole $d_2/\lambda$, while the first is kept on the surface ($d_1 = 0$). The results are shown in Fig.4, which clearly demonstrates the advantages of using high Q modes of larger spheres. Note the resulting negative sign in Eq.(19) due to odd $n$ for $a_{39,1}$ which is why the curves in Figs.3b and 4b are flipped upside down. We have also verified that very similar results are obtained if the dipole's frequency is in the vicinity of another WGM, namely, $a_{2311,167}$. Our numerical results suggest a nice "single mode" approximation

$$\begin{aligned} K_s + i\Omega_s &\approx \kappa K_{0s}\left[\kappa - i(\omega - \omega_G)\right]^{-1}, \\ K_d + i\Omega_d &\approx \pm \kappa K_{0d}\left[\kappa - i(\omega - \omega_G)\right]^{-1}, \end{aligned} \qquad (21)$$

where $\kappa = \omega_G / 2Q$ is the half width of the WGM and the sign + (-) is for the even (odd) mode. $K_{0s}$ and $K_{0d}$ differ from each other by the strengths of the mode functions for the WGM at $\vec{r}_A$ and $\vec{r}_B$.

We have thus shown that the radiative interaction between distant, otherwise noninteracting dipoles can be mediated via the excitation of whispering gallery modes of micro-spheres. For high Q modes and dipole frequency close to the mode frequency, our calculations suggest the applicability of the approximation where the contributions from the surface dependent Green's function can be replaced by one arising only from a single WGM. Our results should be important in several contexts, for example, in the studies of energy transfer between donor-acceptor molecule pairs. The transfer rate can be enhanced by several orders of magnitude, as it is proportional to the square of the dipole-dipole interaction, which, as shown, can reach giant values when mediated by whispering gallery modes of larger spheres. We further note that there exist several possibilities for the experimental realization of the theoretical predictions. One could adsorb the donor-acceptor pair on the surface of the micro-sphere. The other possibility, in the spirit of classic experiments of Ref.9, could be to let the micro-sphere fall between two parallel atomic beams. Finally we note that in the foregoing we have considered the case of weak radiative coupling between the atom and the vacuum of the electromagnetic field. The regime of strong coupling is interesting in its own right and leads to important consequences [10,15] if such interactions are mediated by the WGM's.

One of the authors (SDG) would like to thank the Department of Science and Technology, Government of India for supporting this work.

## APPENDIX A

Quantum electrodynamics of the radiative interaction

In this appendix we present a brief derivation of the radiative interaction. Let $\vec{E}(\vec{r},t)$ represent the quantized electromagnetic field. In dipole approximation the interaction Hamiltonian of the atoms with the field is

$$H_1(t) = -\sum_{\alpha} \vec{d}_\alpha \cdot \vec{E}(\vec{r}_\alpha, t) \left| e_\alpha \right\rangle \left\langle g_\alpha \right| e^{i\omega_0 t} + H.C., \tag{A1}$$

where we use the interaction picture and sum over the two atoms. The second order contribution to the wave function is

$$\left| \Psi^{(2)} \right\rangle = -\frac{1}{\hbar^2} \int_{-\infty}^{t} dt_1 \int_{-\infty}^{t_1} dt_2 H_1(t_1) H_1(t_2) \left| \Psi^{(0)} \right\rangle \tag{A2}$$

The initial state $\left| \Psi^{(0)} \right\rangle$ is $\left| e_A, g_B, 0 \right\rangle$, where $\left| 0 \right\rangle$ represents the vacuum state. The final state is $\left| g_A, e_B, 0 \right\rangle$ and hence the transition amplitude $A$ will be

$$A = -\frac{1}{\hbar^2} \int_{-\infty}^{t} dt_1 \int_{-\infty}^{t_1} dt_2 \left\langle g_{A,} e_B, 0 \left| H_1(t_1) H_1(t_2) \right| e_A, g_B, 0 \right\rangle . \tag{A3}$$

A calculation shows that the rate of change of the transition amplitude in the long time limit is given by

$$\begin{aligned} \frac{\partial A}{\partial t} \to &-\frac{1}{\hbar^2} \int_0^\infty \left\langle \vec{d}_B \cdot \vec{E}(\vec{r}_B, \tau) \vec{d}_A^* \cdot \vec{E}(\vec{r}_A, 0) \right\rangle e^{i\omega_0 \tau} d\tau \\ &-\frac{1}{\hbar^2} \int_0^\infty \left\langle \vec{d}_A^* \cdot \vec{E}(\vec{r}_A, \tau) \vec{d}_B \cdot \vec{E}(\vec{r}_B, 0) \right\rangle e^{-i\omega_0 \tau} d\tau, \end{aligned} \tag{A4}$$

which on using the properties of the electromagnetic field vacuum can be written as

$$\begin{aligned}\frac{\partial A}{\partial t} &= -\frac{1}{\hbar^2}\int_0^\infty dt\, e^{i\omega_0\tau}\left\langle [\vec{d}_B\cdot\vec{E}(\vec{r}_B,t),\vec{d}_A^{\,*}\cdot\vec{E}(\vec{r}_A,0)]\right\rangle \\ &= \frac{i\omega_0^2}{\hbar c^2}\vec{d}_B\cdot\vec{G}(\vec{r}_B,\vec{r}_A,\omega_0)\cdot\vec{d}_A^{\,*}.\end{aligned} \tag{A5}$$

The last line in (A5) follows from the definition of propagators in quantum electrodynamics in terms of the commutators of the electric field operators. Note the close relation of (A5) to Eq.(7), which we derived from classical arguments. The above argument clearly shows the relation of radiative interactions to the propagators. The argument leading to (A5) is quite general. Thus if the medium between the two atoms is something else, then the radiative interaction will be obtained by using the modified propagator.

## Figure Captions

Fig.1 Schematic view of the micro-sphere of radius $a$ with two similarly oriented dipoles on the z-axis, at distances of $d_1$ and $d_2$ from the surface of the micro-sphere. Refractive index of the sphere material is $\sqrt{\varepsilon} = 1.46$, with that of outside medium $\sqrt{\varepsilon_0} = 1.0$.

Fig.2 (a) Extinction coefficient $Q_{ext}$ and (b) dipole-dipole characteristics $K_d$ (solid line) and $\Omega_d$ (dashed line) as functions of detuning $\omega_0 - \omega_G$ for $d/\lambda = 0.1$ for $a_{2312,167}$ mode. Both the curves in (b) are normalized to the enhancement factor $K_{0d} = \max(|K_d|)$ =1349.81. The corresponding curves for $d/\lambda = 1.0$ are identical except that now $K_{0d}$ =154.47.

Fig.3 Same as in Fig.2 except for $a_{39,1}$ mode. The corresponding enhancement factors for distances $d/\lambda$ =0.1 and 1.0 are now 3.4 and 0.07, respectively.

Fig.4 Dipole-dipole enhancement factor $K_{0d}$ in the vicinity of the cavity resonance as a function of $d_2/\lambda$ for (a). $a_{2312,167}$ and for (b) $a_{39,1}$ modes. The first dipole is on the surface. The dashed lines show the same when the resonant mode is suppressed. At $d_2/\lambda = 0$, these values are 0.00394 and 0.070, respectively.

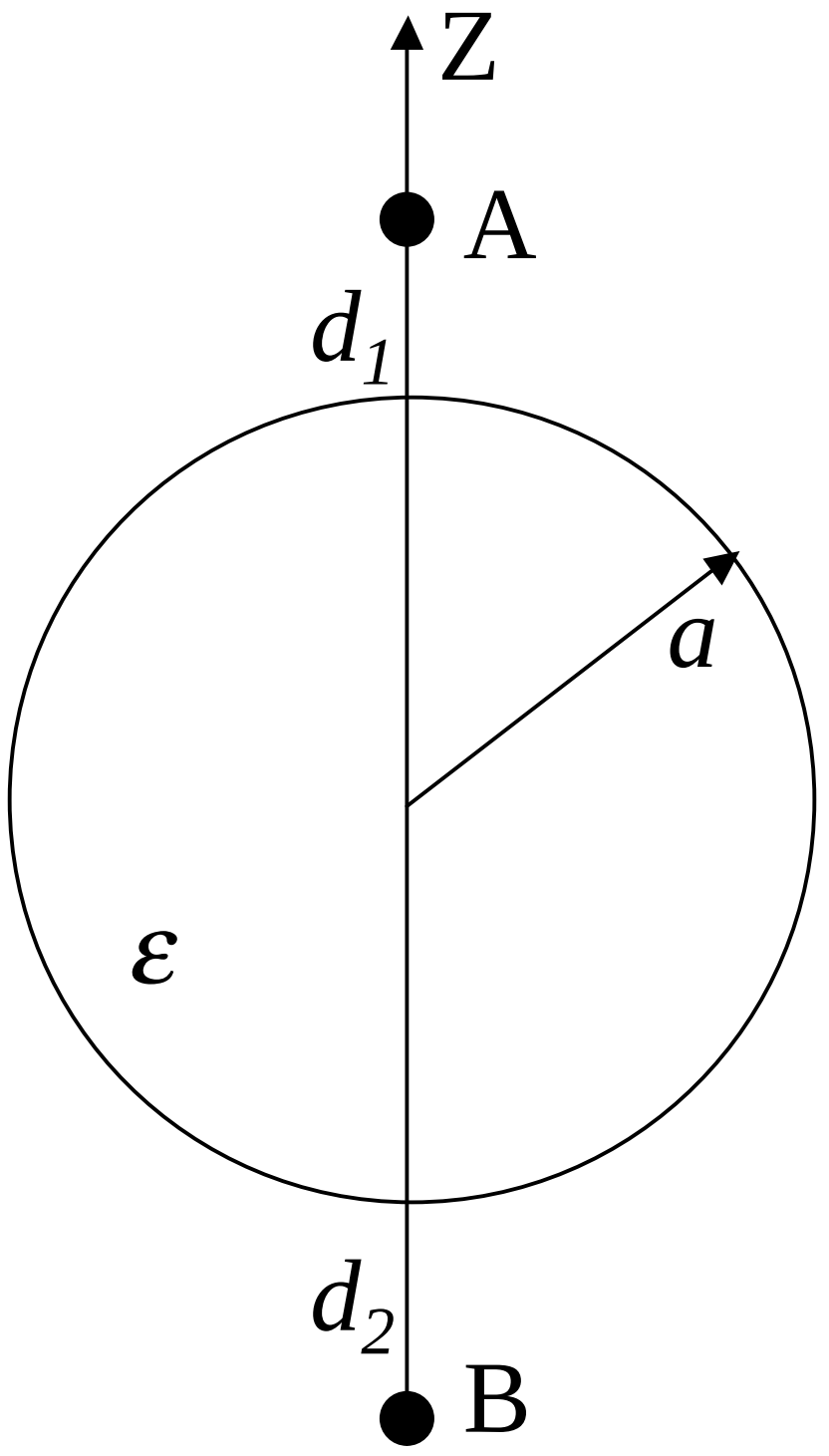

Fig.1 G S Agarwal and S Dutta Gupta,
"Giant radiative…"

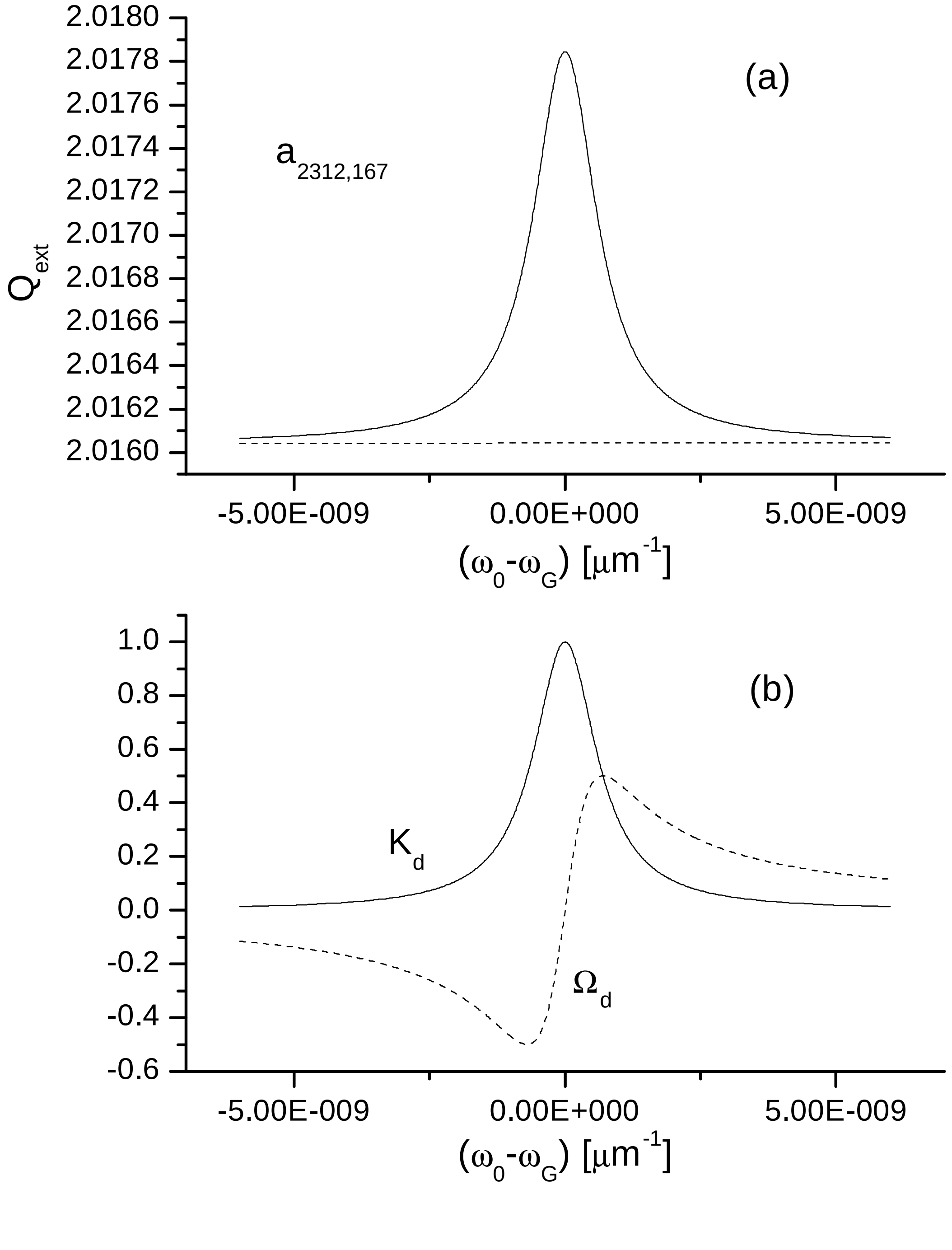

Fig.2. G. S. Agarwal and S Dutta Gupta, “Giant radiative …”

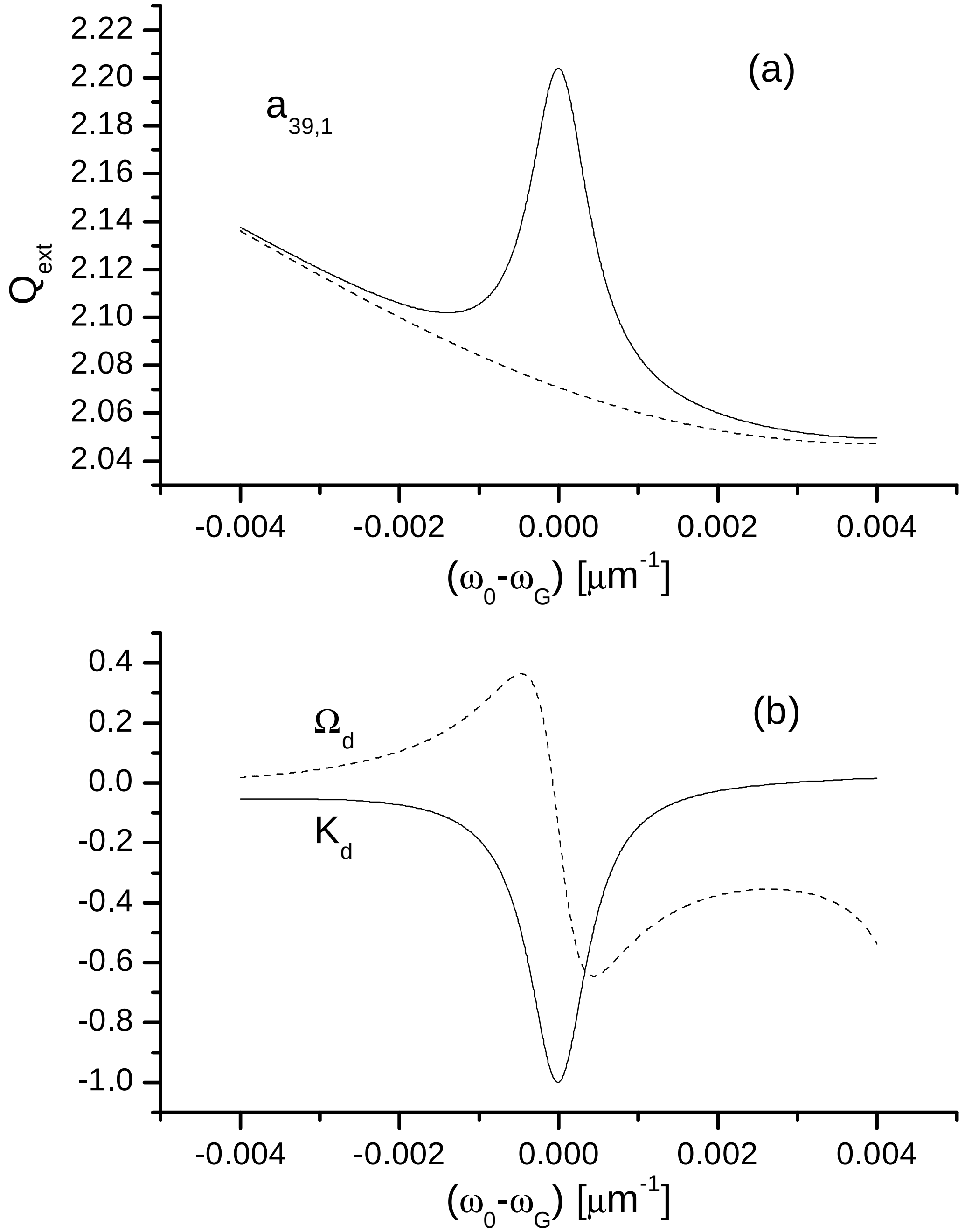

Fig.3. G. S. Agarwal and S Dutta Gupta, "Giant radiative …"

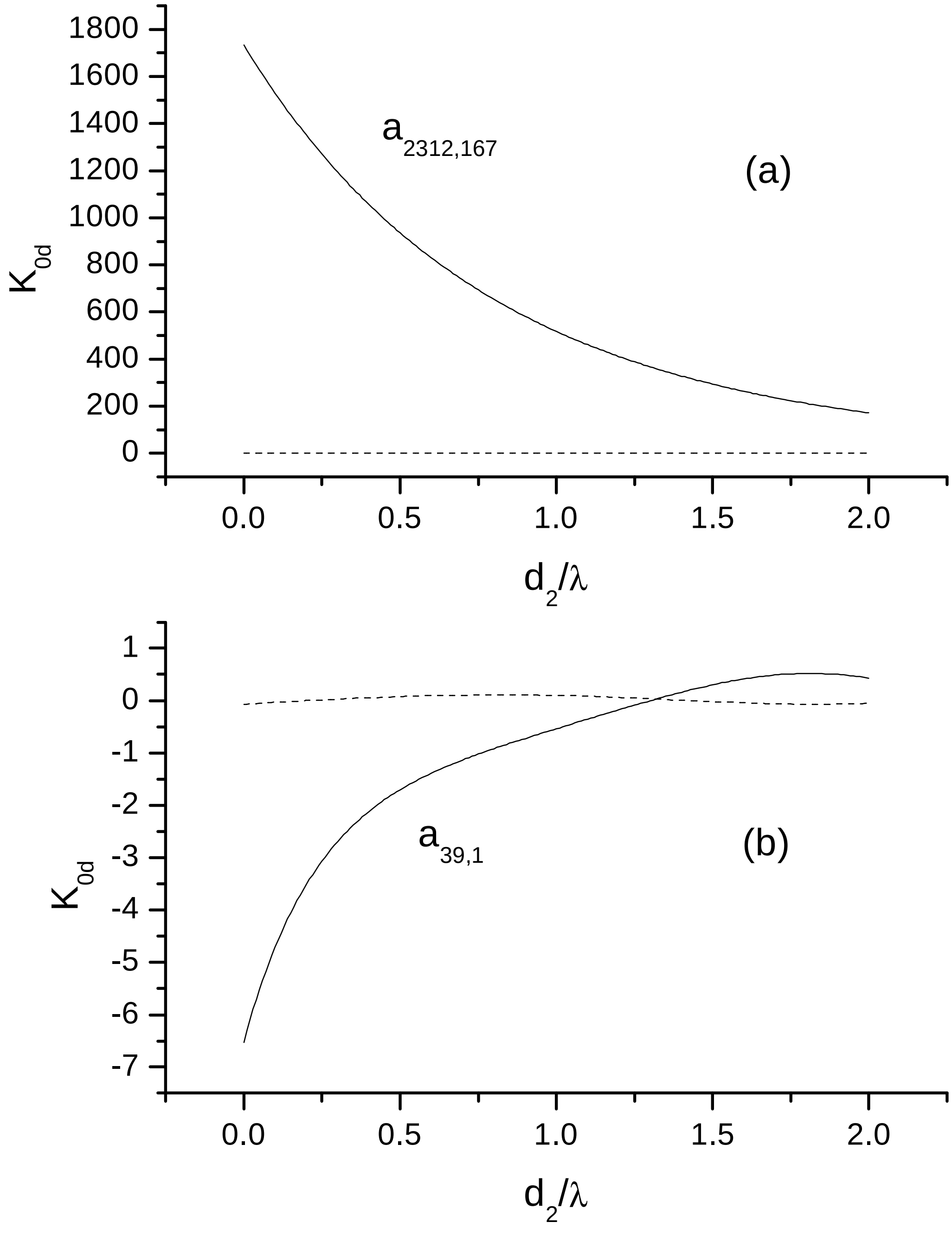

Fig.4. G. S. Agarwal and S Dutta Gupta, “Giant radiative …”